\documentclass[aps,twocolumn]{revtex4} 
\usepackage{graphicx}
\usepackage{amssymb}        
\usepackage{amsmath}  
\usepackage{epsfig}  
\usepackage{color}

\begin{document}

  \title{Reconstruction of eye movements during blinks} 
  \author{M.  S.  Baptista$^{1,4}$}
	\address{$^1$Max-Planck-Institut f\"ur Physik
Komplexer Systeme, N\"othnitzerstr. 38, D-01187 Dresden, Deutschland}
\author{C. Bohn$^{2,3}$}
\address{$^2$ Freie Universit\"at Berlin, Institut 
f\"ur Deutsche und Niederl\"andische Philologie, 
Habelschwerdter Allee 45, 14195 Berlin, Deutschland}
\author{R. Kliegl, R.  Engbert}
\address{$^3$ Universit{\"a}t  Potsdam, Institut f\"ur Psychology, 
Karl-Liebknecht-Str. 24/25, 14476 Potsdam OT Golm, Deutschland}
\author{J. Kurths}
 \address{$^4$Universit{\"a}t Potsdam, Institut f{\"u}r Physik, 
    Am Neuen Palais 10, D-14469 Potsdam, Deutschland}
  \date{\today}

\begin{abstract}
In eye movement research in reading, the amount of data plays a crucial role
for the validation of results. A methodological problem for the analysis of
the eye movement in reading are blinks, when readers close their
eyes. Blinking rate increases with increasing reading time, resulting in high
data losses, especially for older adults or reading impaired subjects. We
present a method, based on the symbolic sequence dynamics of the eye
movements, that reconstructs the horizontal position of the eyes while the
reader blinks. The method makes use of an observed fact that the movements
of the eyes before closing or after opening contain information about the
eyes movements during blinks. Test results indicate that our reconstruction
method is superior to methods that use simpler interpolation approaches. In
addition, analyses of the reconstructed data show no significant deviation
from the usual behavior observed in readers.
\end{abstract}

\maketitle


\section{Introduction}

{\bf Studies of eye movements have revealed enormous insights into the
understanding of cognitive processes during visual search, scene viewing and
reading. The development of high resolution techniques that allow researchers
to record the eye movements online during experiments constitute the base for
the research in this field. Systems range from coil systems,
Electro-Oculo-Graphy EOG systems and video based systems (e.g. Eyelink) to
systems using infrared light to capture the pupils movements, such as the Dual
Purkinje, corneal reflection and pupil boundary systems. All of those systems
require that the reader's eyes are open and closing the eyes inescapably means
loss of data measurement. A typical and natural closure of the eyes is the
blink, a brief closing of the eyelids. Blinks can occur spontaneously, as a
normal periodic closing (e.g. for eye lubrication), reflexively (protective,
e.g. due to an air puff or an object moving towards the eyes) and voluntarily
(e.g. by command)
\cite{guitton:1991}. In this work, we propose 
a method, based on the symbolic sequence dynamics of the eye movements, to
reconstruct the eye positions during blinks in a reading experiment, in which
readers were instructed to read 144 isolated sentences of the Potsdam
Sentence Corpus (PSC). The main assumption behind this approach, for
reconstructing the eyes position during a blink, is that the eyes movements
before and after a blink occurs carry information about how the eyes behave
during a blink.}

In reading research, we distinguish between {\it saccades}, rapid eye
movements of the eyes with high velocities, and {\it fixations}, when our eyes
remain relatively still between the saccades. The function of a saccade during
reading text is to bring a new text section into foveal vision, where visual
acuity is highest.

Two important questions arise when the reader closes his eyes. Did the reader
make a saccade ? If the reader did make a saccade, at what time did the saccade
start? The exact starting time of a saccade is crucial for the detection
of fixations and influences fixation durations. The later the saccade happened
during a blink, the longer will be the detected fixation prior to the
blink. Fixation duration is the primary measure related to cognitive processes during reading comprehension.

One approach to treat the reading data with blinks is to exclude the trials
during which blinks occur. A trial includes the recording of the eye positions while
reading a given sentence. A different approach would be to analyze the
trial up to the point in time when a blink occurred. Both approaches might 
exclude relevant information concerning the particular dynamics of eye movements in reading and  blink phases.
Therefore, it would be of high value to the data analysis in
reading research, especially in data sets of older readers, to
reconstruct a complete sequence of saccades and fixations per trial.

If the reader did not make a saccade while the eyes were closed, then the eyes
have not moved during the blink and their position before and after the
blinking is roughly the same. The problem appears when a saccade happens
during a blink and the challenge is to accurately reconstruct such a saccade.

The first difficulty in treating eye movement data is provided by its clear
non-stationarity characteristic (see Sec. \ref{data}), mainly caused by the
reader's goal to read the sentence which commands the spontaneous dynamics of
the eyes composed by the alternation between (micro)saccades
\cite{engbert:2003} and fixations. In addition, once that the reader decides
to make a saccade, there is a random time-delay until the moment the saccade
is actually realized (see Ref. \cite{kliegl:2006}). So, the movement of the
eyes is not only non-stationary but also contains stochastic components. These
characteristics together unable one to model the eyes positions by the usual
time-delay reconstructing technique \cite{takens}.

The second difficulty is the fact that the eyes' dynamics is the result of a
typical complex system \cite{politi}, a system whose output is produced by the
interaction of many systems, which presumably has a high-dimensional output. In
fact, on the one hand, the long-term evolution of the eyes movements have a
high-dimensional character, which calls for some kind of stochastic model
\cite{paul}. On the other hand, the short-term evolution of the eye 
movements possess a dynamic typical of low-dimensional systems, in
particular a saccade.

In order to resolve the first difficulty, instead of working with the eye
position, we work with the eye velocity (see Sec. \ref{symbolic}). The
velocity is the most simple technique to transform a non-stationary data into
a stationary one. This technique is often used to treat complex data. Among
the many works, this technique was used in the analysis of microsaccades
(rapid eyes movements) \cite{ralf}, stock market fluctuations
\cite{baptista_PHYSICAA2002}, and plasma turbulence \cite{baptista_plasma}. 

To resolve the second difficulty, we make models that consider the dual
low-high-dimensional characteristic of the eye movements. First, we model the
short-term evolution of the eye oscillations, in particular a saccade, using a
low-dimensional system, a one-dimensional damped oscillator (see
Sec. \ref{dynamics}).  The main parameters that adjust this oscillator to
saccades, the period and the damping coefficient, are obtained from both the
low and high dimensional characteristics of the eye movements.
 
For the purpose of finding such parameters we use a technique suitable for
treating complex data, the symbolic dynamics
\cite{politi,kitchens}.  A complex system with infinite number of possible output values, in
particular the eye velocities, is transformed into  a much simpler system by
encoding their velocities using a few number of letters, creating the symbolic
sequences  (see Sec.  \ref{symbolic}).  In  this  work, we  consider  symbolic
sequences composed of 4 letters.

The symbolic dynamics technique has a powerful property that can be explored
to treat systems that possess different dynamics for different time-scales. A
short-length symbolic sequence generated from short-term time intervals
provides an instantaneous visualization of the movement of the eyes, while a
large-length symbolic sequence generated from long-term time intervals might
reveal stochastic properties
\cite{roland1} of the eye movements, averages, and other
quantities. Stochastic properties can also be obtained by producing averages of many short-length symbolic sequences. These properties are
fundamental to the success of our reconstruction method.

In order to know whether and when a saccade happens, during a blink, we
consider symbolic sequences created by encoding the eye velocities before and
after this blink occurs (see Secs. \ref{symbolic} and
\ref{entropia}). To know whether a saccade happens, we 
compare the probabilities of finding certain groups of short-length symbolic
sequences before and after this blink, with the probabilities of finding those
short-length symbolic sequences before and after all saccades observed during
all the trials in an experiment. At this point, we take advantage of the
statistical character of short-length symbolic sequences. To know when a saccade
happens, we search in the large-length symbolic sequences particular repeating
sequences of letters that correspond to the encodings of oscillations that the
eyes make before or after making a saccade.

Notice that an experiment is composed of many trials. In addition, the
dynamics of the eyes is highly influenced by what is being read, and besides
during an experiment the participant makes a series of blinks. As a
consequence, we cannot construct infinity large symbolic sequences, but
symbolic sequences with a finite length smaller than $L$. Further, we
decompose these sequences in even smaller sequences regarded as words.

We define $L$ in the following way: Symbolic sequences of length larger than
$L$ contain roughly the same content of information of symbolic sequences of
length $L$. So, the long-term character of the symbolic sequences is provided
by finite-size symbolic sequences of length not larger than $L$. For our
experiment, $L$ is of the order of 60, which means symbolic sequences composed
of 60 letters, corresponding to a time interval of 120ms (see
Sec. \ref{entropia}).
 
Once we have ensured that a saccade occurs during a blink and we have
obtained an estimation of the time the saccade starts and its period, the
time series is reconstructed by assuming that the position of the eyes before
and after the occurrence of a blink can be optimally connected by a time series generated
by a simple one-dimensional damped oscillator (see Sec. \ref{reconstroi}).

In order to check if our method can really reconstruct saccades during blinks,
we study its performance to reconstruct the position of the eyes during
artificially created blinks. As shown in Sec. \ref{optimal}, our method can
reconstruct saccades much better than if we had made the reconstruction
assuming that a saccade happens in the middle of a blink, having a period
given by the average period of all the saccades observed in all trials.

Finally, we have shown (in Sec. \ref{analyses}) that relevant reading measures
that reflect cognitive processing are not affected by our reconstruction
method, which validates our method from the psychological point of view.

In appendix A, we show the most important parameters, variables, and constants
considered in this work.

\section{Experimental Data}\label{data}

\begin{figure}[!h]
\begin{center}
 \includegraphics*[width=9.0cm,height=8.5cm,viewport=0 0 450 550]{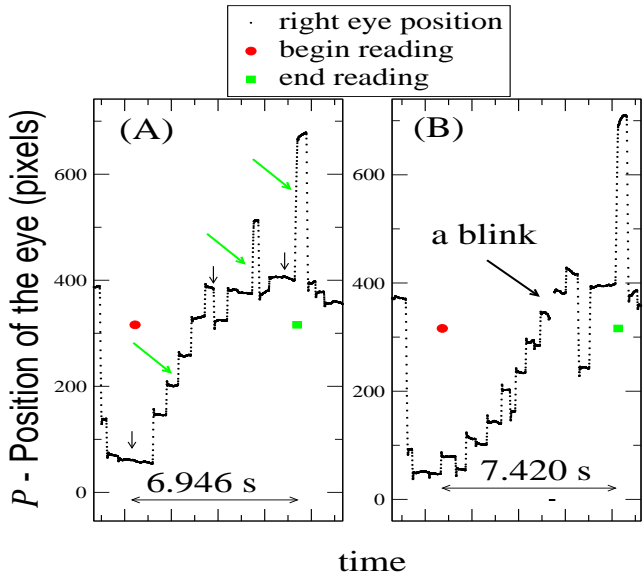}
\end{center}
\caption{[Color online] Points 
represent the horizontal position of the right eye $P$ in pixels in the
computer screen over time, for one trial. The (red) circle indicates the time
at which a sentence is presented on the screen.  The (green) square indicates
the time when the participant finishes reading the given sentence. (A) A few
saccades are indicated by the thick inclined arrows and a few fixations by the
thin vertical arrows. (B) A blink is indicated by the arrow. During the blink,
the participant's eyes made a saccade.}
\label{paper_fig01}
\end{figure}

{\it Data.} In this work, we consider only data from the horizontal position
of the right eye, denoted by $P$. A sample of the raw data set is shown in
Figs. \ref{paper_fig01}(A-B), where $P$ in pixels is plotted over time. In
both figures, the typical alternation between saccades and fixations can be
observed. In Fig. \ref{paper_fig01}(B) we observe a data loss (the position
$P$ is set to zero) due to a blink that occurred while the eyes made a
saccade, indicated by the misalignment of the signal before and after the
blink.
 
{\it Participants}. The method of reconstruction was applied on data sets of
four young participants (mean age 18.5 years) and three old participants (mean age 72.3
years). All participants had normal or corrected to normal vision. 
Visual acuity was assessed with a standard optical chart  (Landolt rings; 5 m distance).
Eye movements of 26 other young participants and 22 other old
participant serve as a baseline to compare gains of the reconstructed data with non
reconstructed data. Each individual data set of the seven participants is
identified by $I_{\eta}$, with $\eta=[1,\ldots,7]$. Average number of blinks per
minute for the seven subjects was 6.97 blinks/min.

{\it Apparatus}. Single sentences were presented on the center line of
a 21-in. EYE-Q 650 Monitor (832 pixels x 632 pixels resolution; frame rate 75
Hz; font: regular, New Courier, 12 point) controlled by an Apple Power
Macintosh G3 computer. Participants were seated in front of the monitor with
the head positioned on a chin rest. Eye movements were recorded with an
EyeLink II system (SR Research, Toronto) with a $\tau$ = 2 ms sampling
rate. All recordings and calibrations were binocular.

{\it Procedure}. In an experimental reading study, participants were instructed to
read 144 isolated sentences of the Potsdam Sentence Corpus (PSC) for
comprehension (cf. Ref. \cite{kliegl:2004} \cite{kliegl:2006} for further
details on material and procedures).

\section{Dynamics of the saccade}\label{dynamics}

\begin{figure}[!h]
\begin{center}
 \includegraphics*[width=8.0cm,viewport=0 -10 450 520]{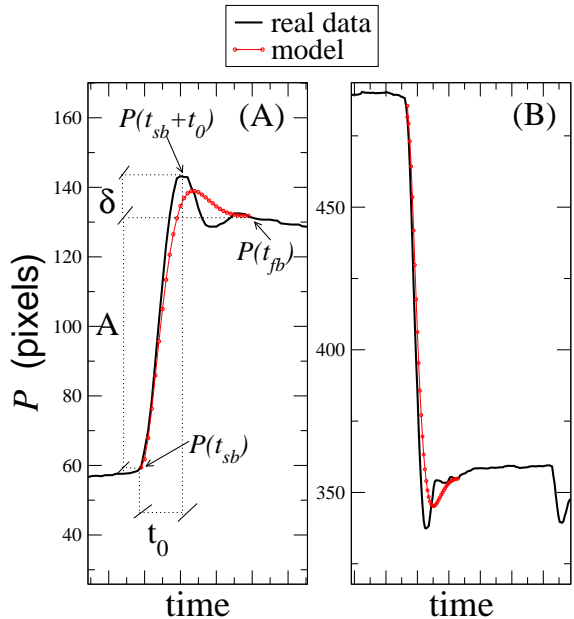}
\end{center}
\caption{[Color online] Thick black line represents a typical rightward (A) and leftward (B)
saccade. Solution of Eq. (\ref{model_saccade}) is shown by the gray (red) line.}
\label{paper_fig04}
\end{figure}

Two typical saccades are shown in Figs. \ref{paper_fig04}(A-B) in black solid
line.  In (A), the position of the right eye moves to the right, making a
rightward saccade, while in (B), the eye moves to the left, making a leftward
saccade. A saccade begins at time $t_{sb}=i \tau$ if $|P({i+1}) - P({i})|>4
pixels$, for $i=[t_{sb}/\tau+k]$ and $k=1,3$. The period of the saccade $t_0$
is determined by checking if $P({m+1}) - P({m})$ changes the signal once, for
$m>t_{sb}/\tau+3$.  So, $t_0=(m-i)\tau$ms. The final position of the eye after
a saccade, the beginning of a fixation happens for the time $t_{fb}$
if for $c>t_{sb}/\tau +t_0$, the signal of $P({c+1}) - P({c})$ changes three
times. While reading all the  sentences a participant makes $N_s$ saccades.  Each
saccade can be identified by the following parameters $[A(j),\delta(j),$
$t_{sb}(j),t_0(j),$ $t_{fb}(j)]$, with $j=[1,N_s]$, where $A(j)+\delta(j)$ is
the amplitude of the saccade $j$, being $A(j)=|P(t_{fb}) - P(t_{sb})|$ and
$\delta(j)=|P(t_{sb}+t_0) - P(t_{fb})|$. These quantities are represented in 
Fig. \ref{paper_fig04}(A).

We model the dynamics of the horizontal position of the eye during a saccade
in the same way eyelid saccades were modeled in Ref.
\cite{malbouisson:2005}, using a damped oscillator described by 

\begin{equation}
  \ddot{X}(t^{\prime}) + 2g(t^{\prime}) \dot{X}(t^{\prime}) + \omega^2X(t^{\prime}) =0 
\label{model_saccade}
\end{equation}
\noindent
with $X(t^{\prime}=0)$=$A(j)$ and $\dot{X}(t^{\prime}=0)$=0.
Equation (\ref{model_saccade}) can describe a saccade in the coordinate
system of the eye position $P(t^{\prime})$, for the saccade $j$, if the following
transformation is applied
\begin{widetext}
\begin{equation}
P(t^{\prime}) = \left\{ 
\begin{array} {r@{\quad:\quad}l} X(t^{\prime}) + 
P(t_{fb}) & P(t_{fb})-P(t_{sb})<0 \\
-X(t^{\prime})+P(t_{sb}) + A(j) & P(t_{fb})-P(t_{sb})>0.
\end{array} \right.
\label{transforma0}
\end{equation}
\end{widetext}
\noindent
The time $t^{\prime}$ in Eq. (\ref{model_saccade}) is transformed
to the time $t$ of the experiment by 
\begin{equation}
t^{\prime} = t-t_{sb}
\label{transforma1}
\end{equation} 

The saccades are modeled by Eq. (\ref{model_saccade}) with parameters $g$
(damping coefficient) and $\omega$ (angular frequency of the saccade
oscillation) such that this equation describes a supercritical damped harmonic
oscillator. This hypothesis is sustained by the fact that most of the time
$\delta(j) \ll A(j)$.  However, in 10$\%$ of the saccades we find
that $\delta(j)
\approx A(j)$ or $\delta(j)$ is slightly bigger than $A(j)$, which
violates this hypothesis. For a more general model for saccades one could use
the dynamical system proposed in Ref. \cite{ackman:2005}. However, for the
scope of the present work, Eq.  (\ref{model_saccade}) is a sufficiently good
model for a saccade.  Note that if $\delta(j) \approx A(j)$ that usually
implies that $A(j)$ is small and therefore, for a short time interval the
saccade can be well approximated by the supercritical model which produces
almost a straight line.  In addition, the effect of this small saccade in the
data treatment is small.  Then, for the saccade $j$, we assume that
\begin{equation}
g(j)=\frac{1}{t_0(j)}\log{\left(\frac{A(j)}{\delta(j)}\right)}
\label{equacao_g}
\end{equation}

and 

\begin{equation}
\omega(j)=\frac{\pi}{t_0(j)}. 
\end{equation}

In conclusion, a saccade can be uniquely defined by the following minimal set
of three parameters $[g(j),\omega(j),A(j)]$.  Using the obtained minimal set
of parameters, we model the saccades of Figs. \ref{paper_fig04}(A-B) using
Eqs.  (\ref{model_saccade}), (\ref{transforma0}), and (\ref{transforma1}). The
model qualitatively reproduces the real data, as can be seen in
Figs. \ref{paper_fig04}(A-B). During a blink, however, in case a saccade
happens, $g$ and $w$ are undetermined and will be estimated by searching for
special patterns in the symbolic encoding of the eye velocities of time series
before and after the occurrence of a blink.

\section{Symbolic characterization of the eye dynamics}\label{symbolic}

One of the most efficient techniques to turn a non-stationary data set into a
stationary set is to work with the velocity space, instead of using the phase
space. The velocity of the eye is given by $V(i
\tau)=[P((i+1)\tau)-P(i\tau)]/\tau$, which will be denoted by $V_i$.  The
velocity is still a very complex variable. To reduce its complexity without
compromising its content of information, we symbolize the data series using 
a small-size alphabet composed of 4 letters: $\{0,1,2,3\}$. 

\begin{figure}[!h]
\begin{center}
 \includegraphics*[width=8.0cm,viewport=0 0 550 500]{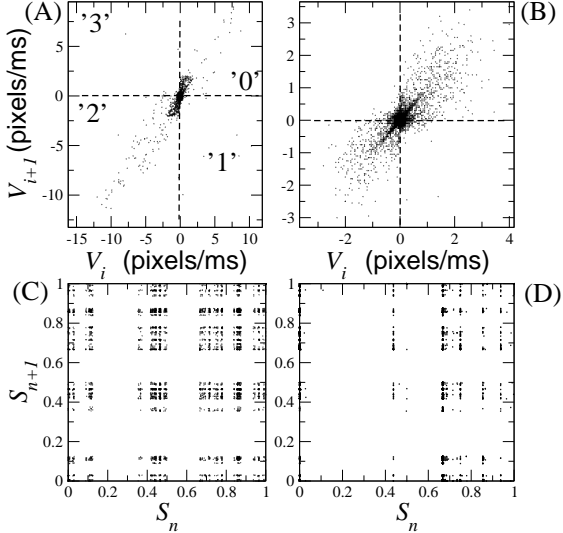}
\end{center}
\caption{The velocity space for all short time series
($t_d \tau=48$ms) before a saccade begins (A), at the time $t_{sb}(j)$, and
after the saccade $j$ reaches its maximum (B), at the time
$t_{sb}(j)+t_0(j)$. Symbolic spaces of the pre-saccade sequences, $s_k(pre)$,
(C) and the post-saccade sequences, $s_k(post)$ (D). Every real in (C-D)
encodes a word of 24 letters, i.e. $t_d$=24.}
\label{paper_fig03}
\end{figure}

The velocity space, a first returning map of the velocity variable, is shown
in Fig. \ref{paper_fig03}(A). Points in the horizontal axis represent the
relation between the velocity at the "time" $i$ [$i \tau$, in units of ms] and
in the vertical axis, the velocity at time $i+1$ [$(i+1)
\tau$]. In this figure, we plot the velocity space for all short time
series ($t_d \tau=48$ms) before a saccade begins, at the time $t_{sb}(j)$. That is,
we consider the time interval $[t_{sb}(j)-t_d \tau, t_{sb}(j)]$. In Fig. 
\ref{paper_fig03}(B), we plot the velocity space for all short time series ($t_d \tau=48$ms)
right after the saccade $j$ reaches its maximum, at the time
$t_{sb}(j)+t_0(j)$. So we consider the time interval $[t_{sb}(j)+t_0(j),
t_{sb}(j)+t_0(j)+t_d \tau]$. Notice that there is a clear visual difference
between both velocity spaces, which reflects a difference in the eye movements
before and after saccades happen.

The symbolic encoding of the velocity variable is done by first splitting the
velocity space in the four partitions as represented by the dashed lines in Fig. \ref{paper_fig03}(A-B),
 and then assigning letters if the velocity trajectory is within
one of these partitions: if $V_i>0$ and $V_{i+1}>0$, we encode this point by a
'0'; if $V_i
\geq 0$ and $V_{i+1} \leq 0$, we encode this point by a '1'; if $V_i <
0$ and $V_{i+1} < 0$, we encode this point by a '2'; if $V_i \leq 0$ and
$V_{i+1} \geq 0$, we encode this point by a '3'. Thus, a trajectory with
$t_d+1$ points ${V_{i},V_{i+1},V_{i+2},\ldots, V_{i+t_d+1}}$ is encoded into a
symbolic sequence with $t_d$ letters, regarded as $s_k(j)$, with
$k$=$[1,t_d]$. $t_d$ is assumed to be an even number, so we can easily split
the symbolic sequences into two non-overlapping sequences.

For the visualization of the symbolic sequences, we split them into two
non-overlapping words of $t_d/2$ letters, and encode these two words into two
real numbers $S_n(j)$ and $S_{n+1}(j)$ by the following rule

\begin{eqnarray}
  S_{n}(j) &=& \sum_{k=1}^{t_d/2} s_k 4^{-k-1} \nonumber \\
  S_{n+1}(j) &=& \sum_{k=t_d/2+1}^{t_d} s_k 4^{-k-t_d/2-1} \label{transforma2}
\end{eqnarray}
\noindent
Then, we plot the numbers $ S_{n}$ versus $S_{n+1}$. 

The existence of blinks together with the fact that sentences have a limited
number of symbolic sequences, and further the fact that an experiment consists of many
trials unable one to consider longer symbolic sequences. Instead, we work
with symbolic sequences of length smaller than
$L$. These symbolic sequences are constructed in a way to
reflect some particular characteristic of the eye movements.

In this work, we consider the following types of symbolic
sequences: The length-$t_d$ symbolic sequences, $t_d \leq L$, $s_k(pre,j)$ and
$s_k(post,j)$ are constructed using the encoding of the eye movements before
the saccade $j$ starts (time interval $[t_{sb}(j)-t_d \tau, t_{sb}(j)]$) and
after this saccade ends (time interval $[t_{sb}(j)+t_0(j),
t_{sb}(j)+t_0(j)+t_d \tau]$), respectively.  The symbolic spaces of $s_k$ are
shown in Fig. \ref{paper_fig03}(C-D), which are the symbolic spaces of the
velocity spaces of Figs.  \ref{paper_fig03}(A-B), respectively. They show 
a clear mismatch between the eye movements before and after saccades happen. 
The length-$t_d$ symbolic sequences, denoted by $s_w(pre,j)$ and
$s_w(post,j)$, are constructed by the encoding of the eyes velocities before
and after a blink that happens at the time $l_i \tau$ms.

Finally, we define the symbolic sequences $s_s(j)$ relative to the saccade $j$
in the following way. For a fixed $t_d$, we construct non-overlapping words of
length $K t_d$, with $K \geq 3$, from the velocity variable, for all valid
trials, i.e. all time intervals of the experiment where blinks are absent and
where there is no interruption of the reading due to the beginning or ending
of the sentences. Then, the symbolic sequences $s_s(j)$ are all symbolic
sequences formed by 3 pairs of length-$t_d/2$ words in the form
$s_{b1}.s_{b2}.s_{b3}.s_{a1}.s_{a2}.s_{a3}$, such that a saccade happens
during the time at which $s_{b3}$ was generated. The symbolic space for
participant $I_{\eta}$, denoted by $\xi(I_{\eta})$, is constructed by
converting the pairs of words $\{s_{b1}.s_{b2}\}$,$\{s_{b2}.s_{b3}\}$,
$\{s_{b3}.s_{a1}\}$, $\{s_{a1}.s_{a2}\}$, and $\{s_{a2}.s_{a3}\}$, into pairs
of real numbers using Eqs. (\ref{transforma2}).  The symbolic space
$\xi_{pre}(I_{\eta})$ represents the pair of words $\{s_{b1}.s_{b2}\}$ and
$\xi_{post}(I_{\eta})$ represents the pair of words $\{s_{a2}.s_{b3}\}$. These
symbolic spaces for a participant are shown in Figs. \ref{paper_fig07}(A-C).

The rules that describe the way short-length words appear after short-length
words in the pre and post saccades symbolic sequences [$s_k(pre)$ and
$s_k(post)$, respectively], also know as the {\it grammar}, is described by
the digrams that give the possible transitions and probability of transitions
$p$ between words of $D$ letters. For a given $D$, with $D << t_d$, we split
the symbolic sequence in a sequence of non-overlapping words of length $D$,
and analyze their probability transitions.  Given a symbolic sequence $s_k$
composed by $k$ letters, with $k=[1,t_d]$, we generate the words, regarded as
$s^{\prime}_q$, each of length $D$. As an example, for $D$=2, given a sequence
$\{s_1\ s_2\ s_3\ s_4\}$ we create the sequence $s^{\prime}_q$=$s^{\prime}_1\
s^{\prime}_2$, with $s^{\prime}_1$=$\{s_1\ s_2\}$ and $s^{\prime}_2$=$\{s_3\
s_4\}$.  The probability of having a length-$D$ word $s^{\prime}_q$ followed
by $s^{\prime}_{q+1}$ is represented by $p_{s_{q}^\prime.s_{q+1}^\prime}$.

\begin{figure}[!h]
\begin{center}
 \includegraphics*[width=8.0cm,viewport=0 0 550 770]{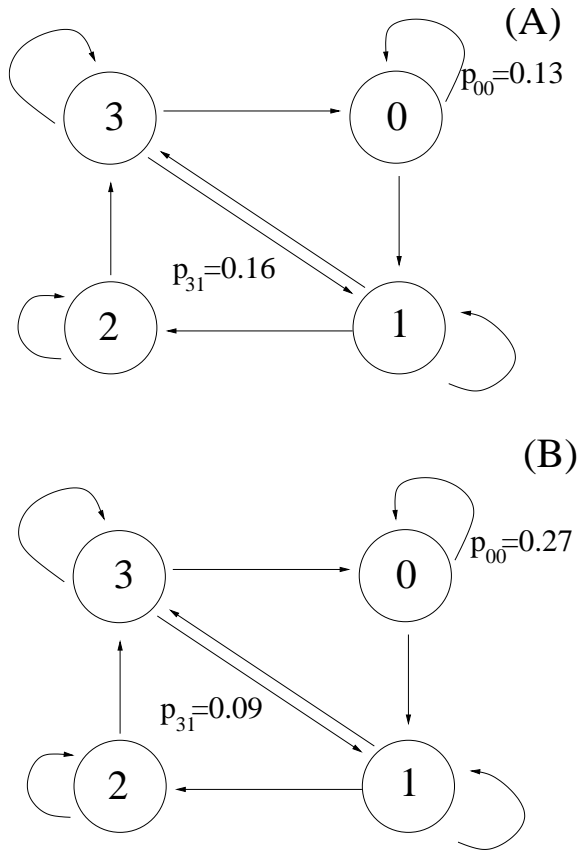}
\end{center}
\caption{Digrams representing the observed transitions among one-letter words
in the 
pre-saccade (A) and post-saccade (B) symbolic sequences of the participant $I_5$.}
\label{paper_fig05}
\end{figure}

Taking participant $I_5$ as an example, for $D$=1, the digram for the one-word
letters in the pre-saccade symbolic sequences, $s_k(pre)$, (with $t_d$=24
symbols) is shown in \ref{paper_fig05}(A), and for the post-saccade symbolic
sequences, $s_k(post)$, in Fig.  \ref{paper_fig05}(B).  We only show the most
different probability transitions.

For example, the probability of finding a '0' that is followed by a '0' in the
pre-saccade symbolic sequences is 0.13 while it is 0.27 in the post-saccade
symbolic sequence. The discrepancy might be a combined effect due to many
factors. The damping of the eye movements after the saccade (after the period
$t_0$) results in smaller amplitudes and slower oscillations in the eye
movements, reflected in symbolic sequences that present more length-2 sequences
of zeros (positive velocity) as well as more length-2 sequences of twos
(negative velocity). Note, however, that in the long term, $s_k(post)$ does
not present large repeating sequences of zeros or twos, because it reflects
the dynamics of the fixations, when the eye randomly oscillates around a fixed
point. This causes the big empty areas in Fig. \ref{paper_fig03}(C), specially
within [0,0.4].

The large probability transition, $p_{31}$, for the pre-saccade sequences is a
consequence of the fact that the pre-saccade velocities change its sign more
often, possibly a consequence of the fact that in the pre-saccade the eyes
present faster oscillations.

It is important to notice that the digrams of length-$D$ words in the pre and
post saccades symbolic sequences do not change significantly by varying $t_d$
in the interval  $40ms \geq t_d
\tau
\geq 120$ms. This means that the digram dynamics presents a sort of
time invariance, if $D$ is kept constant, while varying $t_d$.

In the following, we will consider in our reconstruction method $D$=2, since
for $D>2$ no significant better performance of the reconstruction method was
obtained. 

The digrams represent the statistical character of the eye movements before
and after a saccade happens, provided by averages from the probabilities of
many short-length words observed in all symbolic sequences $s_k$.  They will
be used to identify whether a saccade occurs during a blink.

\section{Predictions based on the symbolic dynamics}\label{entropia}

As a first step, we calculate the Shannon's entropy $H(t_d)$ of the pre- and
post-saccade symbolic sequences, $s_k(pre)$ and $s_k(post)$, of length $t_d$,
in order to estimate an upper bound for the time interval $(L \tau)$ so that predictions can
be made by using the information contained in these symbolic sequences

\begin{equation}
H(t_d) = - \sum_{m} p_m \log_2{(p_m)}
\label{shannon_entropy}
\end{equation}
\noindent
where $p_m$ represents the probability of finding a particular
symbolic sequence $s_k$ composed of $t_d$ letters. 

\begin{figure}[!h]
\begin{center}
 \includegraphics*[width=8.0cm,viewport=0 0 500 500]{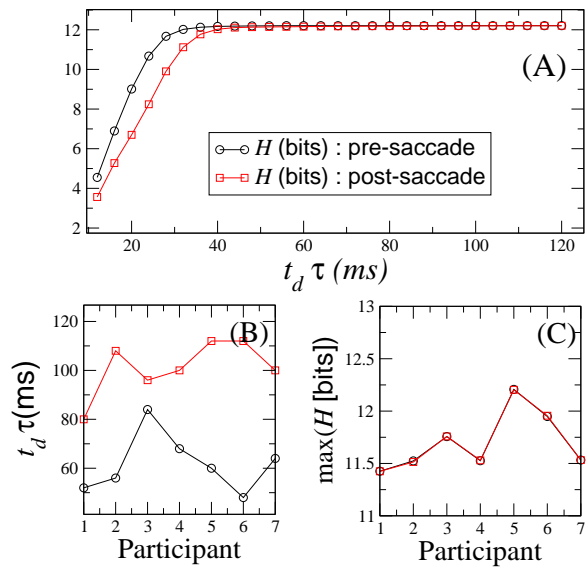}
\end{center}
\caption{[Color online] Circles indicate the analysis in the pre-saccade sequences 
and squares in the post-saccade.  (A) $H$ with respect to $t_d \tau$, for
participant $I_4$. (B) The time interval for which the maximum of the entropy is
reached and in (C) the value of these maximums.}
\label{paper_fig06}
\end{figure}

In Fig. \ref{paper_fig06}(A), we show $H$ for participant $I_4$, which is an
estimation of the amount of information for the symbolic sequences $s_k(pre)$
and $s_k(post)$, for all saccades of the experiment. Empty circles show the
entropy of the sequences $s_k(pre)$, and empty squares show the entropy of the
sequences $s_k(post)$. An important point that can be derived from the shape
of this curve is that the maximum entropy values for both sequences is
approximately equal. The maximum of the entropy is reached
when $\Delta H(t_d) \leq 0.05$, with $\Delta H (t_d)$ defined as 
$$\Delta(t_d) = H(t_d+1) - H(t_d).$$
\noindent
Thus, once the time interval $t_d$ provides sequences for which the maximum of
the entropy is reached, an increase in the the length of the symbolic
sequences does not increase the content of information.

However, these maxima are reached for different $t_d
\tau$.  This becomes clear in Figs.
\ref{paper_fig06}(B-C). In (B), the value of the time interval
$t_d \tau$, when the maximum value is reached for each one of the participants,
is shown. Clearly, the maximum of the entropy is reached for shorter time in
$s_k(pre)$ than in $s_k(post)$. The time intervals in (B) give an estimate of
the time length one can still extract relevant information from the symbolic
sequences, i.e., an estimation of the value of $L$.  The presence of a maximum
for the entropy is a result of either under-sampling or the presence of correlations in the
data set. That also reflects that our data is composed by a finite number of
"repeating" trials that last for a finite time. In (C), the maximum of the entropy is visualized.

As a second step, in order to derive predictions from symbolic sequences, we have to
assume that a certain degree of dynamical constraint exists in the data, and
that words are dynamically connected to words that appear previously. Given for example
 the following symbolic sequence: ``3 3 1 2 3
0 . 0 1 3 1 3 0 . $s_{c3}$ . $s_{o1}$ . 2 2 3 0 0 0 . 1 3 3 1 2 2'', with
$s_{c3}$ and $s_{o1}$ representing two words with 6 letters,  we expect that
$s_{c3}$ can be derived from the two words that precedes it, denoted by
$s_{c1}(j)$="3 3 1 2 3 0" and $s_{c2}(j)$="0 1 3 1 3 0", and $s_{o1}(j)$ can
be predicted from the two words that appear after it, denoted by 
$s_{o2}(j)$="2 2 3 0 0 0" and $s_{o3}(j)$="1 3 3 1 2 2".

The eye movements can be reduced to two main types of behaviors:  Saccades and
fixations. Since fixation is approximately constant and could be modeled by a
zero velocity behavior, for the purpose of the present work, the only
relevant behavior to be predicted in the sequences $s_{c3}(j).s_{o1}(j)$ is the occurrence of a
saccade.

Throughout this paper we make two more fundamental assumptions.

\begin{description}
\item[Assumption I] A saccade is likely to occur
 within the time interval that creates the sequence $s_{c3}(j)$ [$s_{o1}(j)$] if
the digrams of the pre-saccade [post-saccade] symbolic sequences is likely to be a generator for
  the word pairs $s_{c1}(j)$.$s_{c2}(j)$ [$s_{o2}(j)$.$s_{o3}(j)$].
\end{description}

Likelihood to be a generator depends on a defined probabilistic measure. We
define the quantity
\begin{equation}
\Delta p = \sum_{q=1}^{t_d/D} p_{s^{\prime}_q.s^{\prime}_{q+1}} 
\label{prob_dist}
\end{equation}
\noindent
In this equation, $s^{\prime}_q.s^{\prime}_{q+1}$ represents a pair of
length-2 words ($D$=2) observed either in $s_{c1}(j)$.$s_{c2}(j)$ or
$s_{o2}(j)$.$s_{o3}(j)$. The probability of finding $s_{q}^{\prime}$ followed
by $s_{q+1}^{\prime}$ in the digrams of either the pre or post-saccade
symbolic sequences is denoted by $p_{s^{\prime}_q.s^{\prime}_{q+1}}$.  That
generates the quantities $\Delta p(pre) $ and $\Delta p(post)$, respectively.

A saccade is likely to have happened during the time at which $s_{c3}(j)$ was
generated if $\Delta p(pre) > \Delta p(post)$.  The idea here is that if
$\Delta p(pre)>\Delta p(post)$ in the word pair $s_{c1}(j).s_{c2}(j)$, the
digram that represents the pre-saccade dynamics likely describes the dynamics
that generated this pair. That is, a saccade is likely to have happened
during the time at which $s_{c3}(j)$ was generated. So, Eq. (\ref{prob_dist}) is
a probabilistic distance of how the dynamics of the pre or post-saccade (given
by the digrams) conforms with either the word pairs $s_{c1}(j).s_{c2}(j)$ or 
$s_{o2}(j).s_{o3}(j)$. 

The occurrence of a saccade time is consider to be undetermined if one of the
following conditions apply: (i) for at least one sequence
${s^{\prime}_q.s^{\prime}_{q+1}}$ in both $s_{c1}(j).s_{c2}(j)$ and
$s_{o2}(j).s_{o3}(j)$, we find that $p_{s^{\prime}_q.s^{\prime}_{q+1}}(pre)=0$
and $p_{s^{\prime}_q.s^{\prime}_{q+1}}(post)=0$; (ii) defining the
probabilistic distance between the digrams of the pre-saccade and post-saccade
sequences as
\begin{equation}
\Delta d = \sum_q |p_{s^{\prime}_q.s^{\prime}_{q+1}}(pre)-
p_{s^{\prime}_q.s^{\prime}_{q+1}}(post)|, 
\label{distance_UD}
\end{equation}
\noindent
we find that $|\Delta p(pre) -  \Delta p(post)| < \Delta d$.

\begin{figure}[!h]
\begin{center}
 \includegraphics*[width=8.0cm,viewport=0 0 550 500]{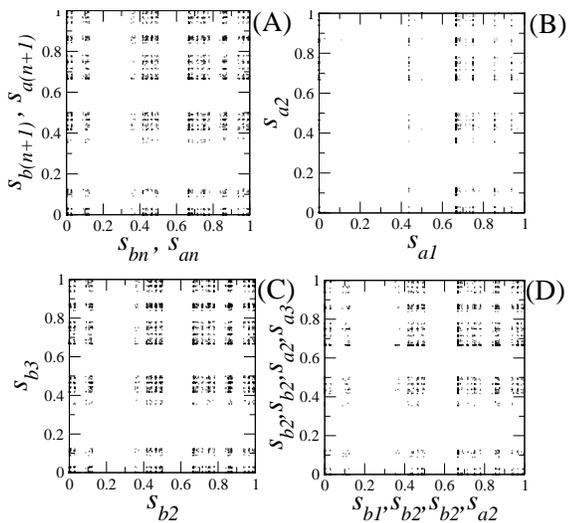}
\end{center}
\caption{The symbolic spaces, $\xi$ in (A), $\xi_{post}$ in (B), 
$\xi_{pre}$ in (C), and  $\xi_{rec}$ in (D), a
transformation of the spaces $\xi_{post}$ and $\xi_{pre}$ that presumably
reproduces the characteristics of $\xi$. These symbolic spaces structures
remain roughly invariant as we consider pre and post-saccade sequences of
length varying from $40ms \leq t_d \leq 120$ms. This points to that the
symbolic dynamics is roughly time invariant. For these figures, we consider
words of length $t_d = 20$, so, as example, $s_{b1}(j)$ represents a 20-letters word. The presence of
large sequences of zeros or twos in $\xi_{pre}$ (C) is responsible for the
points close to 0 and close to 0.7.}
\label{paper_fig07}
\end{figure}

\begin{description}
\item[Assumption II] $\xi(I_{\eta})$ can be approximately obtained by 
applying a transformation {\bf $F$} into $\xi_{pre}(I_{\eta})$ and
$\xi_{post}(I_{\eta})$. We denote the resulting reconstructed space by
$\xi_{rec}$.
\end{description}

\begin{figure}[!h]
\begin{center}
 \includegraphics*[width=8cm,viewport=0 0 500 500]{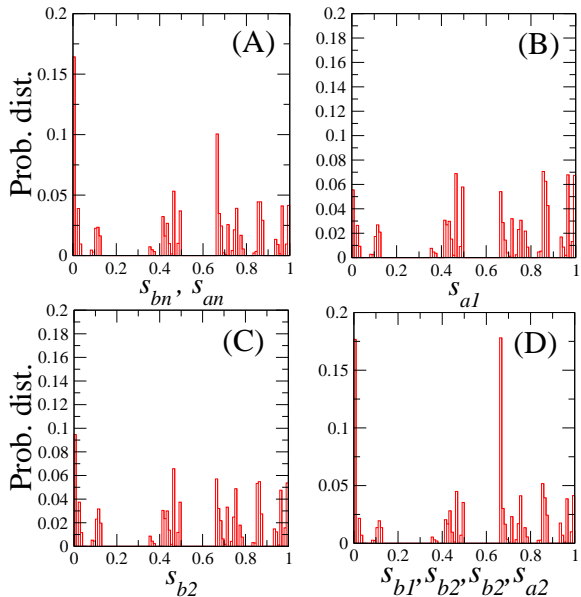}
\end{center}
\caption{The probability distribution of the points in the horizontal 
coordinate of the symbolic spaces in Figs. \ref{paper_fig07}(A-D).}
\label{paper_fig08}
\end{figure}

The result of the transformation {\bf $F$} is shown in
Fig. \ref{paper_fig07}(D), where we plot the symbolic sequences of the form
$s_{b1}(j).s_{b2}(j).s_{b2}(j).s_{a2}(j).s_{a2}(j).s_{a3}(j)$, which produces
the following pair of points $\{s_{b1}(j).s_{b2}(j)\}$,
$\{s_{b2}(j).s_{b2}(j)\}$, $\{s_{b2}(j).s_{a2}(j)\}$,
$\{s_{a2}(j).s_{a2}(j)\}$, $\{s_{a2}(j).s_{a3}(j)\}$. Therefore, the
transformation $F$ results in $s_{b3}$=$s_{b2}$ and $s_{a1}$=$s_{a2}$.  Note
the similarity between the space $\xi$ in (A) and the space $\xi_{rec}$ in
(D). This can also be seen in Fig. \ref{paper_fig08}, where we show the
probability distribution of points in the horizontal coordinates of the spaces
in Figs. \ref{paper_fig07}.

\begin{figure}[!h]
\begin{center}
 \includegraphics*[width=8cm,viewport=0 0 500 500]{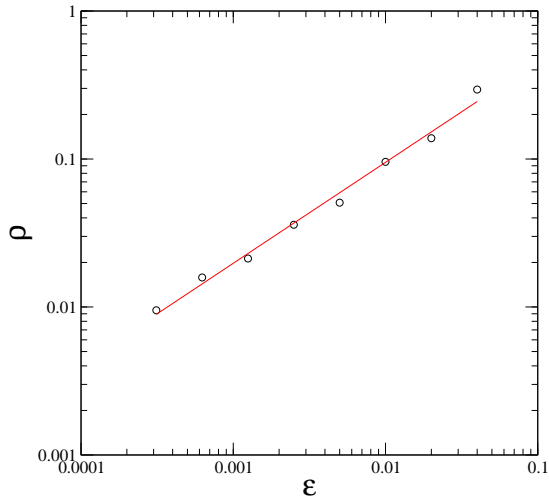}
\end{center}
\caption{Probability of finding in the sequences $s_s(j)$ similar symbolic sequences that respects the 
chosen $F$ transformation.}
\label{paper_fig09}
\end{figure}

The chosen $F$ transformation is one of many other possible dynamics from
which one can successfully recover an equivalent $\xi$ space. But, we choose
it because symbolic sequences of the desired form
$\{s_{b1}.s_{b2}.s_{b2}.s_{a2}.s_{a2}.s_{a3}\}$ are often observed in some of
the $j$ sequences $s_s(j)$. These symbolic sequences are encoded into real
sequences [using Eq. (\ref{transforma2})] into $S_{b1}.S_{b2}.S_{b2}.
S_{a2}.S_{a2}.S_{a3}$. Then, we search sequences of pairs of words in the
symbolic sequence of the saccade, $s_s$, whose real encoding sequence regarded
as $s_r$ is denoted by
$S_{b1}$.$S_{b2}$.$S_{b2}^{\prime}$.$S_{a2}^{\prime}$.$S_{a2}$.$S_{a3}$, with
$S_{b2}^{\prime}$=$S_{b2}+\epsilon$ and
$S_{a2}^{\prime}$=$S_{a2}+\epsilon$. We find that the percentage of points
$\rho$, that respects this rule follows a power-law with respect to
$\epsilon$, $\rho \propto \epsilon^{0.68}$, as shown in Fig.
\ref{paper_fig09}, which implies that $\rho>\epsilon$. 

Therefore, even if we decrease $\epsilon$, there is still a finite
probability, larger than if the relation between $\rho$ and $\epsilon$ were
linear, of finding sequences as the ones created by the transformation $F$.
So, symbolic sequences similar to the desired one appear often, which also
means that points in the symbolic space $\xi(I_{\eta})$ return often to the
diagonal.

One should expect that $\rho \propto \epsilon$ if the symbolic sequences were
generated by a Markov Process. If the probability of
having a symbol '0' is $p_0$ and of having a '1' is $p_1$, the probability of
having the symbol '0' followed by '1' in a Markov process is given by $p_0 \times p_1$. In fact, the
found power-law is an effect of the existence of certain constraints in the
eye's movements, a typical characteristic of dynamical systems. 

The interpretation of the chosen transformation $F$ is simple: The dynamics of
the eyes for short-time intervals before and after making a saccade reveals if
a saccade will happen or has already happened. 

The main difference between the symbolic space $\xi_{pre}$ and $\xi_{post}$ is
the absence (in $\xi_{post}$) of large sequential sequences of zeros or twos,
as the ones observed in $\xi(pre)$. That is easy to be understood by
considering that similarly to what happens in the space for the symbolic
sequences $s_k(post)$ [see Fig. \ref{paper_fig03}(D)], $\xi_{post}$ represents
the long term behavior of the eyes during a fixation, which have roughly a
constant velocity, and does not present tendencies which would be encoded by
larger sequences of zeros or twos.

Another point that is relevant to be emphasized is that before a saccade
happens our eyes have an excited dynamics ($g$ in Eq. (\ref{equacao_g}) is
negative), while after the saccade happens the eyes have a dissipative
(damped) dynamics ($g$ in Eq. (\ref{equacao_g}) is positive). The damping
effect is also another reason for the absence of relevant tendencial
movements.

Thus, $\xi_{post}$ can be considered to represent the autonomous dynamics of the
eyes, while $\xi_{pre}$ represents the forced dynamics. The force is produced by the
will of the participant to move its eyes. Recall that the saccades are reconstructed only considering the dissipative
dynamics in the super-critical behavior.

\section{Approaches for reconstructing the blinks}\label{approaches}

A participant makes $N_B$ blinks during the reading and a
blink occurs at the time $l_i$ (in units of $\tau$, or $l_i \tau$ ms), lasting
$B(l_i)$ (in units of $\tau$ or $B(l_i) \tau$ ms), with $i=1,\ldots,N_B$. As
introduced in Sec. \ref{symbolic}, we
represent the length-$t_d$ symbolic sequences that appear before the
participant blinks by a pair of length $t_d/2$ words $\{s_{c1}.s_{c2}\}$, and
the length-$t_d$ symbolic sequences that appear after the participant opens the eyes is
represented by $\{s_{o2}.s_{o3}\}$. Based on assumption II, the reconstructed symbolic dynamics of the blink is given
by $\{s_{c2}.s_{o2}\}$.

Then we adopt a series of procedures. First, we check whether
$\{s_{c1}.s_{c2}\}$ ($\{s_{o2}.s_{o3}\}$) is likely to be generated by the
digrams of the pre-saccade (post-saccade). In other words, we detect
equivalence of $\{s_{c1}.s_{c2}\}$ with the pre-saccade digram (assumption I).
This suggests that a saccade might have happened right after the participant
blinks, which would imply that in $s_{c2}$ one would find sequences of either
zeros ('0'), if the eyes move to the right after blinking or twos ('2'), if
the eyes move to the left.

If we find a word composed of at least 4 \cite{comment4} sequential zeros
followed one after the other (or twos) in $s_{c2}$, we assume a saccade must have happened
in the time $t_s(l_i)=(l_i+ \theta(l_i))
\tau$ ms, where $\theta(l_i)$ is a function of the number of letters  that precedes the
largest sequential sequence of zeros (or twos) in $s_{c2}$, denoted by
$\theta^{\prime}$. Similarly, we assume that the period of the reconstructed
saccade in units of $\tau$, denoted by $q(l_i)$, is proportional to the number
of sequential zeros (or twos), denoted by $q(l_i)^{\prime}$.

If we detect equivalence of $\{s_{o2}.s_{o3}\}$ with the digram of the
post-saccade, we assume that the saccade must have started in the time
$t_s(l_i)=(l_i+t_d/2+\theta(l_i))\tau$ms, where $\theta(l_i)$ now depends on
the number of letters that appear before the sequential sequences of zeros (or
twos) in $s_{o2}$.  The number of sequential zeros or twos gives the value
$q(l_i)^{\prime}$ from which one can calculate the period of the saccade.

In the case a saccade is likely to have happened during a blink, but
no sequence of only zeros or twos are found either in $s_{c2}$ or $s_{o2}$
that have at least length 4, then, we look for sequences in the whole
reconstructed sequence $\{s_{c2}.s_{o2}\}$. We also search for a sequence of
zeros or twos in the whole reconstructed sequence if the occurrence of a
saccade is undetermined.

In the following section we show how to choose the value of $t_d$ in order to
improve the performance of our reconstruction method. However, while $t_d$ is
considered to be fixed, the $l_i$-th blink has a period $B(l_i)$ which is, in
general, different than $t_d$. If the word pairs $\{s_{c1}.s_{c2}\}$ and
$\{s_{o2}.s_{o3}\}$ show evidence that a saccade happens during the blink,
from the previous considerations this blink will be reconstructed by assuming
that the encoding symbolic sequence of the blink is given by
$\{s_{c2}.s_{o2}\}$. Saccades are reconstructed by finding the repeating
sequences of zeros or twos in $s_{c2}$, $s_{o2}$, or
$\{s_{c2}.s_{o2}\}$. However, the time interval of the blink ($B(l_i)$) is
different than the time interval associated with $\{s_{c2}.s_{o2}\}$ ($t_d$) as
well as the half-time interval of the blink $(B(l_i)/2)$ is different than the
time interval associated with either $s_{c2}$ or $s_{o2}$ ($t_d/2$).

Therefore, if we find that the period of the saccade is $q(l_i)^{\prime}$ (in
units of $\tau$), we rescale this time interval with respect to the time
interval of the blink, and consider that the saccade during the blink has a
period given by
\begin{equation}
  q(l_i)=q(l_i)^{\prime}\frac{B(l_i)}{t_d} \label{equacao10}
\end{equation}
Similarly, the time at which the saccade is assumed to have started is
calculated from $\theta(l_i)$ which is given by
\begin{equation}
  \theta(l_i)=\theta(l_i)^{\prime}\frac{B(l_i)}{t_d} \label{equacao11}
\end{equation}
\noindent
Equations (\ref{equacao10}) and (\ref{equacao11}) are constructed under a
reasonable hypothesis that the time scale of the eye movements during a 
blink are influenced by the period of the blink. They also 
adjust the time interval of the reconstructed symbolic sequences to correspond
to the period of the blink. 

If no words composed of at least 4 zeros (or twos) are found in $s_{c2}.s_{o2}$,
 we assume that a saccade happens in the middle of the blink and the average period $\langle
t_0 \rangle$ of all saccades during reading is taken as the
period of the saccade. If Eqs. (\ref{equacao10}) and (\ref{equacao11}) results in
$q(l_i)+\theta(l_i)>B(l_i)$, we make $\theta(l_i)=B(l_i)-q(l_i)$.
  
Once we have worked out the time that the reconstructed saccade starts, $t_s(l_i)$, and
the period of the saccade $q(l_i) \tau$, we assume in Eq. (\ref{model_saccade})
that $X(t^{\prime}=0)=A(l_i)$, where $A(l_i)$=$P_o(l_i)-P_c(l_i)$, with
$P_o(l_i)$ representing the position of the eye after opening (in the end of
the $l_i$-th blink) and $P_c(l_i)$ representing the position of the eye right
before closing. $t_{sb}$, in Eq. (\ref{transforma1}), should be substituted by $l_i
\tau$. We also assume that after a time given by $l_i \tau + 2.5q(l_i) \tau$, the
reconstructed position of the eyes from Eq. (\ref{transforma0}) is a value so that
$|P(l_i)-P_o(l_i)|$=1 pixel. This means that after 2.5 oscillation periods,
the reconstructed saccade is supposed to end. From this hypothesis, we can
calculate the quantity $\delta(l_i)$, which is used to calculate $g(l_i)$, in
Eq. (\ref{equacao_g}).

A blink $l_i$ is reconstructed by making $P(t)=P_c(l_i \tau)$ for the time
$l_i \tau
\leq t < l_i \tau + t_s(l_i)$ (i.e., no movement of the eyes) 
and for $l_i\tau+t_s(l_i) \leq t \leq l_i\tau + B(l_i)$, we use Eq.
(\ref{model_saccade}) in the coordinate system of the eye position provided by
Eq. (\ref{transforma0}).

\section{Optimal parameters, robustness, and statistical significance of the
reconstruction method}\label{optimal}

To validate our method, we reconstruct blinks artificially
created around the places where saccades happen. So, for each one of the $N_s$
saccades made while reading, we create a series of $f$=10 blinks that last for
a time interval $P_{B}$. The first artificial blink ends at the time $t_{sb}(j)+t_0(j)/2$ ms
(middle of the saccade) and the last artificial blink begins at this same time
$t_{sb}(j)+t_0(j)/2$ ms. Thus, each artificial blink, referred to as $b_r(j)$, with
$r=\{1,\ldots,f\}$ starts at the time $t_{ab}(j,r)$:
\begin{equation}
t_{ab}(j,r) = t_{sb}(j) + t_0(j)/2 - P_{B} + r\frac{P_{B}}{f}. 
\end{equation}

The reconstruction of the artificial blink uses our method previously
described (Secs. \ref{dynamics}, \ref{symbolic}, \ref{entropia}, and 
\ref{approaches}). We first
constructed the digrams, using the symbolic sequences $s_k$ of length $t_d$,
generated from the whole data series, without the presence of artificial
blinks. Then, for every artificial blink we generate the symbolic sequences
$s_w(pre) = \{s_{c1}.s_{c2}\}$ and $s_w(post) = \{s_{o2}.s_{o3}\}$ as
described in Sec. \ref{symbolic} and \ref{approaches}.

After applying our method, for every artificial blink $r$ around a saccade
$j$, we find that the reconstructed saccade starts at the time $t_r(j,r)$ (in
units of ms) and has a period $t_{0r}(j,r)$ (in units of ms).

\begin{figure}[!h]
\begin{center}
 \includegraphics*[width=9cm,viewport=0 0 500 500]{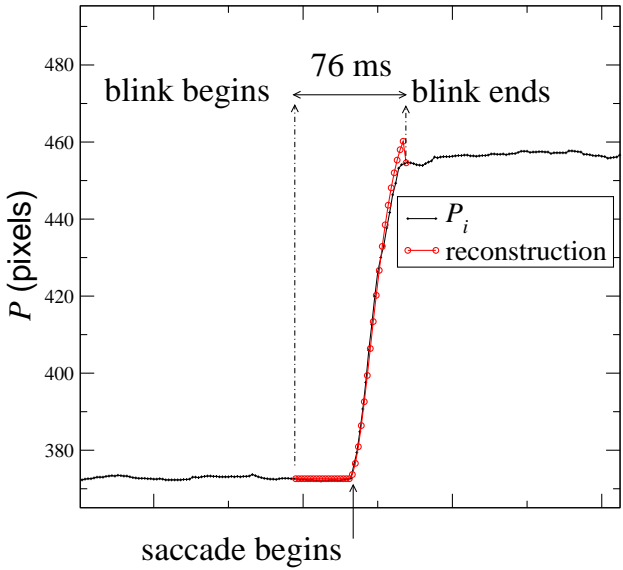}
\end{center}
\caption{[Color online] With empty circles we show the reconstruction of an artificial blink.}
\label{paper_fig10}
\end{figure}

An example of the reconstruction of an artificial blink for the participant
$I_2$ is shown in Fig. \ref{paper_fig10}, indicated by the empty circles. For
this saccade, we have detected that the symbolic sequence of the post-blink,
$s_w(post)$, was likely generated by the digram of the
post-saccade. Remarkably, indeed, the saccade happens right before the end of
the artificial blink. Notice that after the artificial blink ends, one notes a
slight movement of the eyes to the right (increasing the value of $P$), caused
by the typical oscillatory behavior after the saccade happens.  This
oscillatory behavior produces in the symbolic sequence, $s_w(post)$, the
sequential sequences of zeros and twos from which we obtain the time at which
the reconstructed saccade begins [$t_r(j,r)$] and the period of the
reconstructed saccade [$t_{0r}(j,r)$] within an artificial blink.

To quantify how good is the reconstruction of the artificial blink, we
calculate the absolute difference between the predicted time for the beginning
of the saccade, $t_r(j,r)$, and the real time for the beginning of the
saccade, $t_{sb}(j)$, normalized by the time interval of the artificial
blink. We calculate this quantity for all the artificial blinks and average it, producing
\begin{equation}
  \langle t_r \rangle = \frac{1}{N_s.f} \sum_{j=1}^{N_s} \sum_{r=1}^f 
  \frac{|t_r(j,r)- t_{sb}(j)|}{P_{B}}
\end{equation}

Then, we calculate the absolute difference between the real time
when the saccade starts, $t_{sb}(j)$, and the time of a hypothetical reconstructed
saccade if it had started in the middle of the artificial blink, normalized by
the assumed time interval of the artifical blink:
\begin{equation}
  \langle t_r^{\prime} \rangle = \frac{1}{N_s f} \sum_{j=1}^{N_s} \sum_{r=1}^f 
  \frac{|t_r^{\prime}(j,r)- t_{sb}(j)|}{P_{B}}
\end{equation}
\noindent
where $t_r^{\prime}(j,r)$=$t_{ab}(j,r)$+$P_{B}/2$.

Similarly, we calculate the absolute difference between the period of the
reconstructed saccade, $t_{0r}(j,r)$, and the real period
$t_0(j)$, normalized by the average period of the saccades, producing
\begin{equation}
  \langle t_{0r} \rangle = \frac{1}{N_s f}  \sum_{j=1}^{N_s} \sum_{r=1}^f 
  \frac{|t_{0r}(j,r) - t_0(j)|}{\langle t_0 \rangle}
\end{equation}
\noindent
and we calculate the absolute difference between the real period of
the saccade, $t_0(j)$, and the period of a hypothetical reconstructed
saccade if it had been reconstructed assuming that it is equal to the
average period of the saccades $\langle t_0 \rangle$, producing
\begin{equation}
  \langle t_{0r}^{\prime} \rangle = \frac{1}{N_s f} \sum_{j=1}^{N_s} 
\sum_{r=1}^f 
  \frac{|\langle  t_0 \rangle - t_0(j)|}{\langle t_0 \rangle}
\end{equation}

Our method produces statistically significant results if the quantities
\begin{equation}
    \sigma= \langle t_r^{\prime} \rangle - 
    \langle t_r \rangle
\end{equation}
\noindent
and
\begin{equation}
    \beta= \langle t_{0r}^{\prime} \rangle - 
    \langle t_{0r} \rangle
\end{equation}
\noindent
are positive. Notice that a positive $\sigma$ is a measure of 'how much better
our method reconstructs the time a saccade happens' if compared to a method
that would reconstruct a saccade as if it had happened in the middle of the
artificial blink. A positive $\beta$ is a
measure of 'how much better the method reconstructs the period of the saccade'.

\begin{figure}[!h]
\begin{center}
 \includegraphics*[width=8cm,viewport=0 0 500 500]{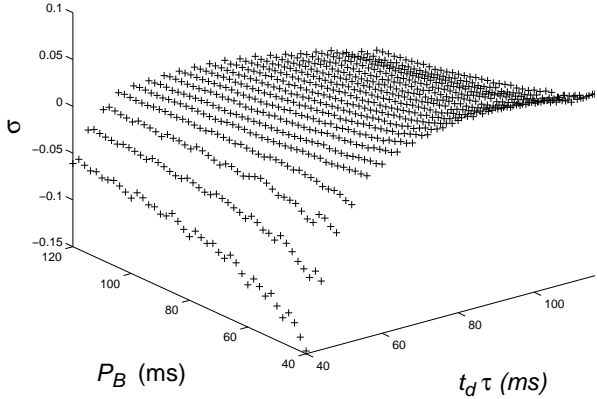}
\end{center}
\caption{Values of $\sigma$ for participant $I_5$
varying $t_d \tau$ and $P_B$, for $D$=2.}
\label{paper_fig33A}
\end{figure}

\begin{figure}[!h]
\begin{center}
 \includegraphics*[width=8cm,viewport=0 0 500 500]{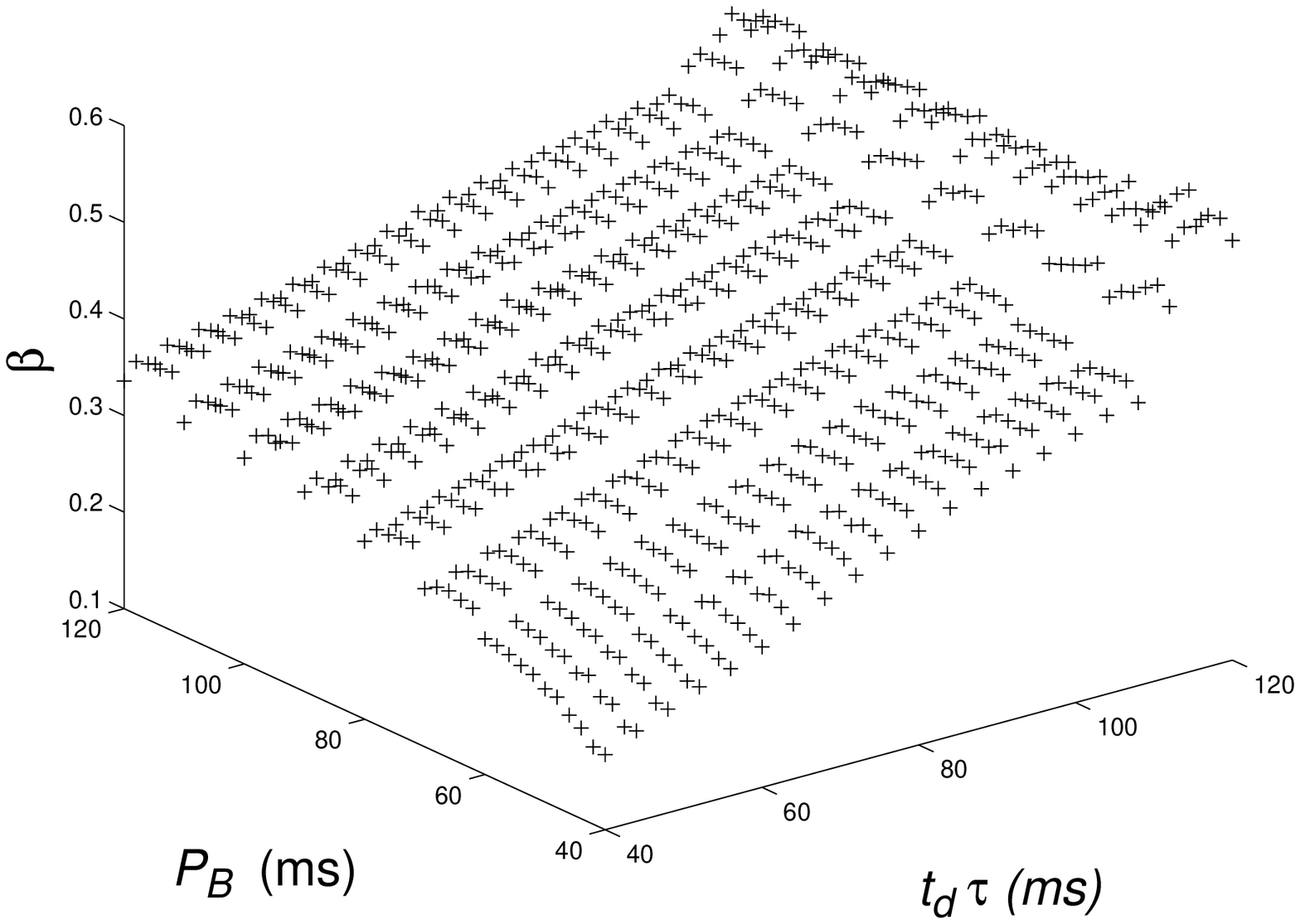}
\end{center}
\caption{Values of $\beta$ for participant $I_5$
varying $t_d \tau$ and $P_B$, for $D$=2.}
\label{paper_fig33B}
\end{figure}

\begin{figure}[!h]
\begin{center}
 \includegraphics*[width=8cm,viewport=0 0 500 500]{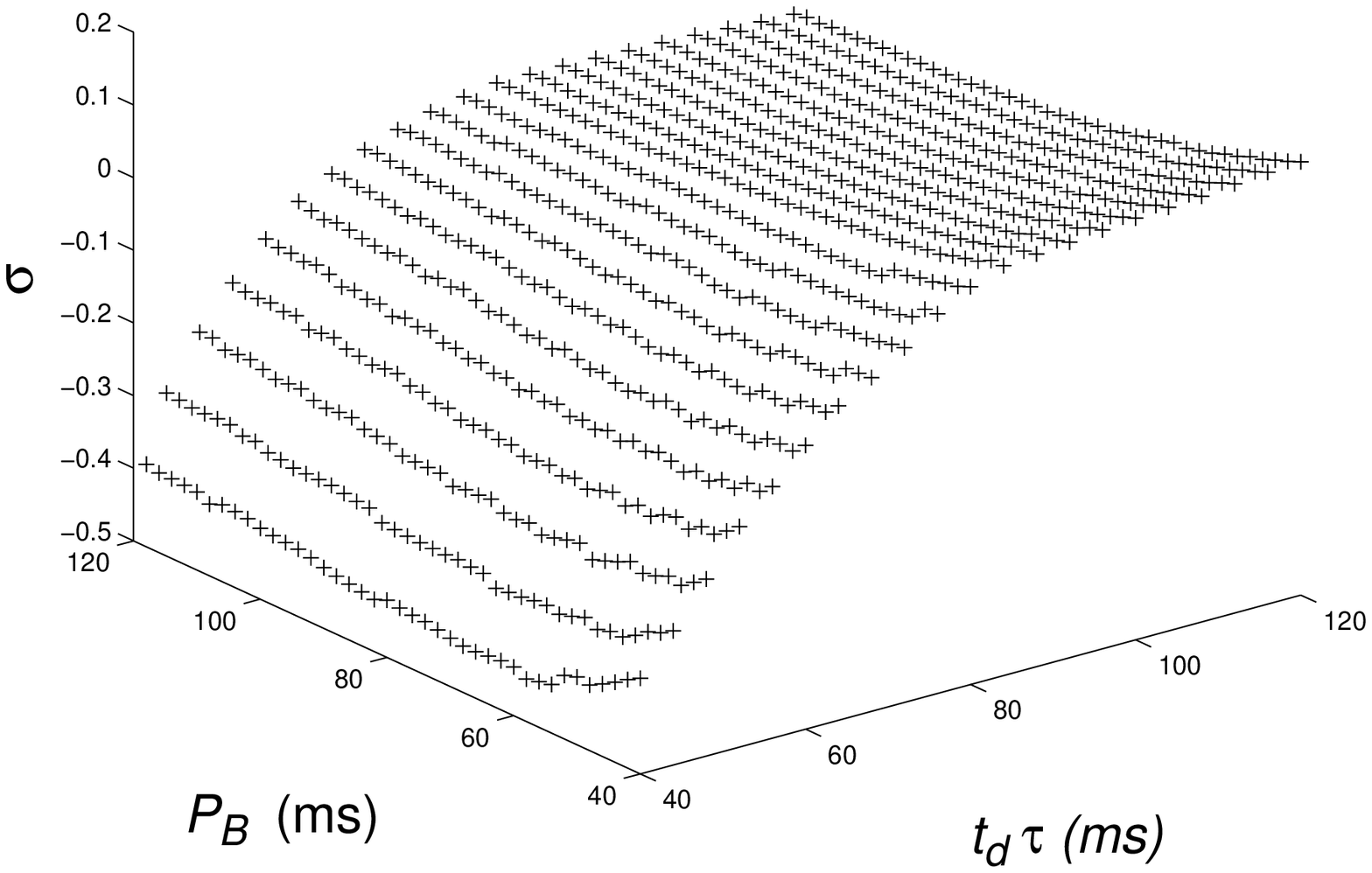}
\end{center}
\caption{Values of $\sigma$ for participant $I_1$
varying $t_d \tau$ and $P_B$, for $D$=2.}
\label{paper_fig06A}
\end{figure}

\begin{figure}[!h]
\begin{center}
 \includegraphics*[width=8cm,viewport=0 0 500 500]{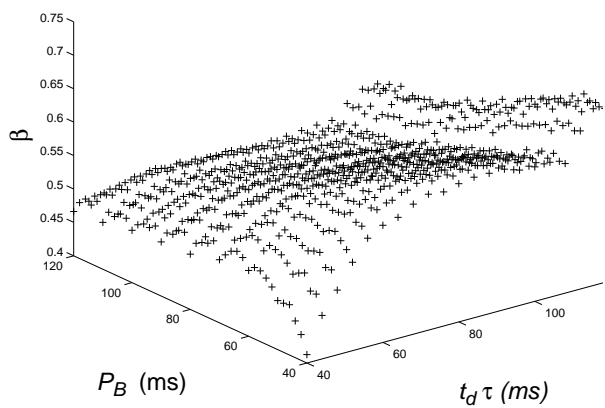}
\end{center}
\caption{Values of $\beta$ for participant $I_1$
varying $t_d \tau$ and $P_B$, for $D$=2.}
\label{paper_fig06B}
\end{figure}

We calculate these quantities by considering different values of the time
interval $P_B$ considered for the artifical blink. For each value of $P_B$, we
reconstruct the artificial blinks using symbolic sequences of different
lengths $t_d$.  In Figs. \ref{paper_fig33A} and
\ref{paper_fig33B}, we show $\sigma$ and $\beta$, respectively, for an young
participant. In Fig. \ref{paper_fig06A} and
\ref{paper_fig06B}, we show the same quantities 
for an old participant.

These figures illustrate that there is an extended region in the parameter
space $t_d
\tau \times P_B$ where parameters can be changed and still our method
reconstructs the saccades present in the artificial blinks in a better way
than if we had just guessed that the saccade starts in the middle of the
artificial blink, having a period equal to the average period of all the
saccades while reading.  This demonstrates the  robustness of prediction of this method,
under a wide range of $P_B$ and $t_d$ values. Similar positive results for the
blink reconstruction were obtained for the other participants.

The quantities $\sigma$ and $\beta$ depend more on the length of the
symbolic sequences, $t_d$, than on $P_B$. This indicates that our method is able to reconstruct real blinks, 
whose durations extend over a wide
range, well enough, if $t_d$ is well chosen. Roughly, $t_d \geq 40$ should produce good
results.

In table 1, for each participant  the average duration of the blinks
$\langle B \rangle$ is listed in the second column, the average period of the
saccades $\langle t_0 \rangle$ in the third column . The value of $t_d$ that
produces the largest value of $\sigma$ and $\beta$ are shown in the columns 4
and 5, respectively. We consider $t_d \leq 120$ms,  a time interval
for which we can extract relevant information from the symbolic sequences (see
Sec. \ref{entropia}). Thus, a reconstructed artificial blink that reproduces
 well the time of the beginning of the saccade is obtained by using $t_d \tau$
shown in the fourth column.  On the other hand, a reconstructed artificial
blink that reproduces well the period of the saccade is obtained by using
$t_d \tau$ values shown in the fifth column. This shows a kind of Heisenberg's
principle in our method. Either one reconstructs reasonably the time a saccade
starts or its period, but not both quantities simultaneously, except for the
readers $I_1$ and $I_3$.

Other quantities that characterize the participants are also shown in this
table. These are the probabilities (column 6) of finding that the pre-blink
symbolic sequences, $s_w(pre)$, (using a $t_d$ value from the fourth column)
were likely generated by the digram of the pre-saccades dynamics, the number
of blinks per second (column 7), the average angular frequency with which the
reader makes a saccade $\langle
\omega \rangle$ (column 8), the average damping coefficient $\langle g \rangle$
of the saccades (column 9), and the time that the participants took to read all
the sentences of the PSC, the quantity $T$(column 10). 

It is interesting to note that participant $I_1$ has a high probability
(0.737) of making saccades immediately after closing the eyes. Furthermore, the time he
takes to read all the sentences of the PSC is the shortest compared to the
other participants. This suggests some sort of efficient reading behavior 
in the sense that the saccade target is already detected before the reader blinks.

Note that for many participants the ideal value of $t_d \tau$ that better
reconstructs the time at which a saccade starts is of the order of the time
interval at which we can still extract relevant information from the symbolic
sequences (see Sec. \ref{entropia}).

\begin{widetext}
\begin{center}
{\small
\begin{table}\label{tabela1}
\caption{Characteristics of the 7 participants considered in this work.}
\begin{tabular}{|c|c|c|c|c|c|c|c|c|c|}
  $I_i$ & $\langle B \rangle$(ms) & $\langle t_0 \rangle$ (ms)& 
$t_d \tau$ (ms) & $t_d \tau$ (ms) & pre-blink & 
$\#$ blinks/s & $\langle \omega \rangle $ & $ \langle g \rangle$ & T (s) \\
  $I_1$   & 60.000  & 41.485 & 120 & 116 & 0.737 & 0.134  & 0.133  & 0.104 & 788.928  \\
  $I_2$   & 75.766  & 35.991 & 76  & 116 & 0.333 & 0.134  & 0.102  & 0.080 & 1149.148 \\
  $I_3$   & 48.424  & 40.045 & 120 & 116 & 0.060 & 0.030  & 0.137  & 0.107 & 1110.560 \\
  $I_4$   & 53.289  & 33.198 & 72  & 116 & 0.000 & 0.046  & 0.118  & 0.093 & 968.500  \\
  $I_5$   & 98.280  & 26.034 & 76  & 116 & 0.236 & 0.098  & 0.185  & 0.099 & 1021.836 \\
  $I_6$   & 72.502  & 23.890 & 76  & 116 & 0.115 & 0.344  & 0.177  & 0.101 & 954.918  \\
  $I_7$   & 72.545  & 23.905 & 92  & 116 & 0.125 & 0.027  & 0.184  & 0.113 & 817.594 
\end{tabular}
\end{table}
}
\end{center}
\end{widetext}

\section{Reconstructing the blinks}\label{reconstroi}

\begin{figure}[!h]
  \centerline{\hbox{\psfig{file=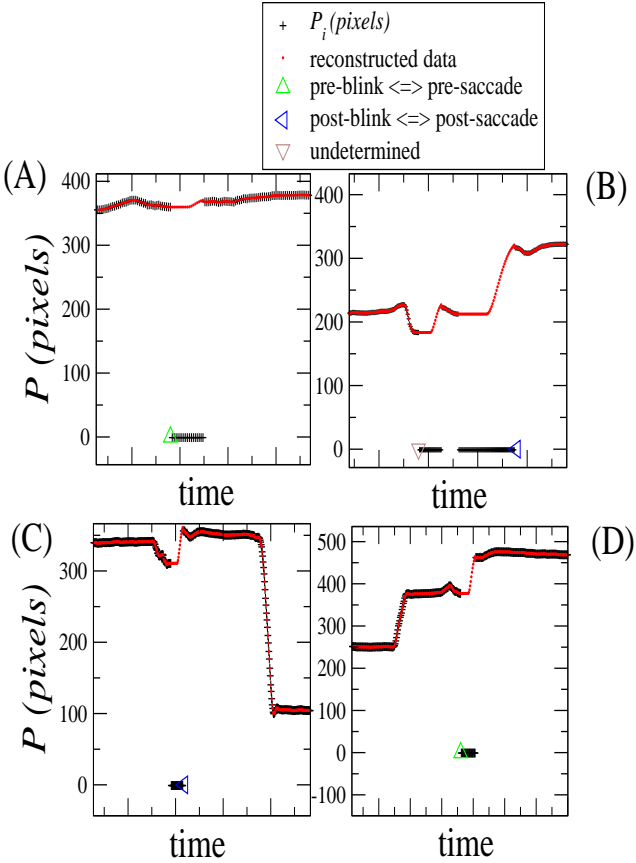,width=9cm,height=12cm}}}
\caption{[Color online] Some examples of the blink reconstruction using our 
method with parameters $D$=2 and $t_d \tau$ from the column 4, in Table 1.}
\label{paper_fig11}
\end{figure}

For reconstructing the real blinks, we consider $D$=2 and $t_d \tau$ from the
column 4 of Table 1, which might allow our method to 
reconstruct well the time at which a saccade starts. We consider symbolic sequences
$s_w(pre)=\{s_{c1}.s_{c2}\}$ and $s_w(post)=\{s_{o2}.s_{o3}\}$, that encode the
eye velocity before and after a blink, respectively. In
Figs. \ref{paper_fig11}(A-D), we show a few examples of the blink
reconstruction for the participant $I_2$. In (A), $s_w(pre)$ was found to be
likely generated by the digram of the pre-saccade symbolic sequence.  In (B),
two blink reconstructions are shown.  In the first blink, the method is unable
to determine if a saccade happens during this blink and it is assumed that a
blink happens in the middle of the interval.  In the second blink, $s_w(post)$
was found to be likely generated by the digram of the post-saccade, which
means that a saccade must have started right before the eyes open, the same
case is illustrated in (C). In (D), $s_w(pre)$ was found to be likely
generated by the digram of the pre-saccade.

\section{Analyses of the reconstructed data}\label{analyses}

The results of the statistical tests provided in Sec. \ref{optimal} and the
reconstruction examples in Sec. \ref{reconstroi} indicate that the model
reconstructs the data losses due to blinks with qualitatively high precision,
 and the reconstructed data seem to reproduce the natural
movement of the eyes.  Of further interest is the benefit we gain from the
reconstruction procedure for the analysis of eye movements in reading.

We have analyzed the reconstructed data using several eye movement measures,
that are considered to be related to cognitive processing during reading, when
the word is the unit of analysis
\cite{rayner:1998}. Gaze duration is defined as the sum of all fixations on a word in first pass reading.
 Total reading time is the sum of
all fixation durations on a word, including regressions from second pass
reading. First pass single fixation duration describes the fixation duration
when a word is fixated exactly once, preceeded and followed by a forward
saccade. If a word is fixated once, fixation position in the word is usually
in the first half of the word, that is on the second or third letter of a word
\cite{oregan:1987}. The skipping rate is the probability a word is not fixated
during the first inspection of the sentence, whereas regression probability
is here defined as the chance that a word is the origin of a regressive eye
movement. All of these measures are important variables if the ease of
processing, strategies during reading, or effects of the material on
fixational behavior (e.g. word frequency effects) are of theoretical interest.

\begin{figure}[!h]
\centerline{\hbox{\psfig{file=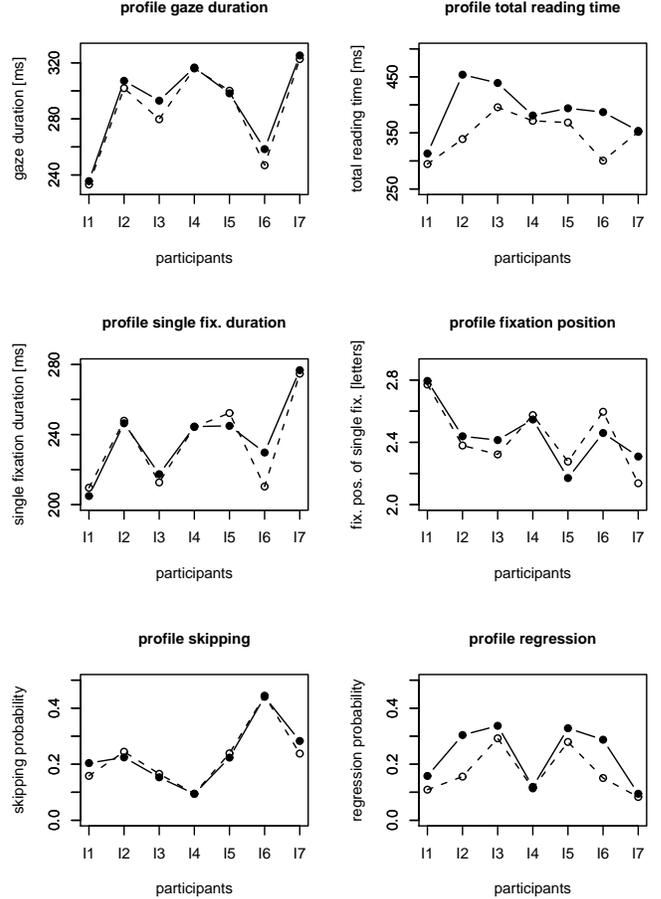,width=9cm}}} 
\caption{Comparison of global eye movement statistics for young and 
old participants as a result of the reconstruction method. Dotted lines
represent results of the unmodeled data, solid lines represent results of the
reconstructed data.}
\label{plot_emid}
\end{figure}

In Figure \ref{plot_emid} means in total reading time per word, gaze
duration, first pass single fixation duration, fixation position of single
fixations, skipping, and regression probability for the data of the seven
participants with and without reconstruction are visualized. Despite
individual differences in eye movement patterns, there are no significant
changes between the original data set and the reconstructed data for any of
the six processing measures. In fact, the qualitative similarity of the data
with and without reconstruction in the various measures of reading behavior is
quite impressive.

Participant $I_2$ and $I_6$ show the largest differences in total reading time
and the probability of regressions. Since the probability of blinking
increases with reading time, i.e. second pass reading, and trials with blinks
were excluded from the analysis in the data without reconstruction, it is
plausible that the valid trials of these two subjects in the data without
reconstruction mostly include trials with few second pass fixations. Total
reading time and the number of regressions depend on second pass reading; in
fact, total reading time correlates with the number of regressions, since
second pass fixation durations are included in total reading time. Therefore,
longer total reading times and higher regression probabilities for these two
subjects in the data with reconstruction is not surprising, but highly
expectable.
 
\section{Conclusions}

We have used the symbolic representation of the eye velocities as a tool to
reconstruct an eye movement signal obtained in a reading experiment, when the
eyes are closed during a blink. The dynamics of saccades played a crucial role for
predicting the signal of the eye positions during blinking. Essentially, for
each participant the signal could be restored by finding the closest
similarities between the dynamics of the eye velocities before or after a
saccade and before or after a blink. Thus, the eye movements before and after
a blink occurs carry information about how the eye behaves during a blink.
Results demonstrated a qualitatively precise prediction for the signals during
blinks.

The reconstructed data did not affect eye movement measures related to
cognitive processing during reading. Total reading time, gaze duration, and
single fixation duration for words did not change significantly, neither was
fixation position in the word, skipping probability or regression probability
affected by the reconstruction of the signal. This was true for young as well
as old subjects. Especially for the analysis of processing measure including
second pass reading (e.g. total reading time, regression probability), the
reconstruction method is a valuable tool to increase statistical power. Blinks
occur more frequently if reading time increases. Thus, trials containing
second pass reading are relatively more often excluded from the analysis, if
blink occurrence is the criteria for trial exclusion. The use of a
reconstruction method recovers those trials with relatively many blinks,
providing a more complete picture of the processing variances between subjects
during reading. The reconstruction of saccades during spontaneous blinks from saccadic data recorded from
the open eye is a useful approximation. We cannot rule out that the blink
itself interferes with the motor program of the saccade, changing its
parameters in subtle ways. As far as we know, at present there is no reliable
evidence in support of this possibility. In sum, the described reconstruction model is a precise and
useful tool to overcome problems of data loss in eye movements during reading,
especially if the analyses of fixation sequences is fundamental to theoretical
questions.

The method proposed here was applied to reconstruct the data series of the
eye position. However, it is of general appliance for other types of complex
data. In a general way, the proposed method is based on the symbolic
identification of either (or both) precursors or indications of the occurrence of an event.
 In the particular case of this work, the event is a saccade. But, in
a general situation, an event could be a heart attack or a brain stroke.  In
many situations, the identification of a precursor of such events would be
desirable in order to take preventive actions \cite{kantz}. But as a first step, one often
wants to know if  an event has happened. Other examples where our method
could provide relevant applications are in the study of extreme events as
earthquakes, stock market crashes, hurricanes, floods, and others.

\textbf{Acknowledgment} This research was supported by Deutsche Forschungsgemeinschaft (grant KL
955/6). MSB was partially funded by the Alexander von Humboldt foundation and
the Helmholtz Center for the Study of Mind and Brain Dynamics at the University
of Potsdam.

\newpage

\begin{widetext}

\section{Appendix}\label{apendiceA}

Definitions of parameters, variables, and constants
\begin{center}
{\small
\begin{tabular}{c|c}
\hline
Symbol & Meaning \\
\hline
\hline
$\tau$ & time step of the experiment ($\tau$=2ms) \\
$P_i$ & horizontal position of the right eye for the time $i\tau$ \\
$I_{\eta}$ & participants $\eta = [1,7]$ \\
$j$ & index used to denote a particular saccade \\
$t_{sb}(j)$ & time that saccade $j$ begins, in units of ms \\
$t_{fb}(j)$ & time that saccade $j$ ends, in units of ms \\
$N_S$ & number of saccades made to read all the sentences \\
$t_0(j)$ & period of the saccade $j$, in units of ms \\
$A(j)+\delta(j)$ & amplitude of the saccade $j$ \\
$\omega(j)$ & angular frequency of the saccade $j$ \\
$g(j)$ & damping coefficient of the saccade $j$ \\
$B(l_i)$ & period of the $l_i$ blink, $i=1,\ldots,N_B$, in units of $\tau$ \\
$N_{B}$ & number of blinks ($l=1,\ldots,N_B$) \\
$ l_i \tau $ & time that a blink happens \\
$V_i$ & velocity \\
$t_d \tau$ & time interval considered to construct a symbolic sequence \\
$t_d$ & length (i.e. $\#$ of letters) of the symbolic sequences considered \\
$S_n$ & real number that represents a symbolic sequence of length $t_d$ \\
$s_k(pre,j)$ & a symbolic sequence of length $t_d$ that ends when the saccade
$j$ starts \\
$s_k(pos,j)$ & a symbolic sequence of length $t_d$ that starts when the saccade
$j$ ends \\
$s_w(pre,l_i)$ & symbolic sequence of length $t_d$ before blink that occurs at
$l_i \tau$ \\
$s_w(post,l_i)$ & symbolic sequence of length $t_d$ after blink that occurs at
$l_i \tau$ \\
$P_m(j)$ & probability of finding a sequence of length $t_d$ \\
$\Delta p$ & summation of the probabilities among transitions of 
length-$D$ symbolic sequences \\
$ \xi  $ & symbolic plane of the saccade \\
$ \xi_{post}$ & symbolic plane of the post-saccade \\
$ \xi_{pre }$ & symbolic plane of the pre-saccade \\
$q(l_i)$ & period of the reconstructed saccade within blink $l_i$, in units of
$\tau$\\
$[l_i + \theta(l_i)]\tau$ & moment that the reconstructed saccade starts
within blink $l_i$, in units of ms\\
$t_s(l_i)$ & time that the reconstructed saccade begins within blink $l_i$, in
units of ms\\
$P_o(l_i)$ & position of the right eye right after the end of the blink
(opening of eye) \\
$P_c(l_i)$ & position of the right eye right before the beginning of the blink
(closing the of the eye)\\
$A(l_i) + \delta(l_i)$ & amplitude of the reconstructed saccade \\
$P_{B}(j,t)$ & time interval of the artificial blink, in units of ms \\
$b_r(r=1,f)$ & artificial blinks around a saccade \\
$t_{ab}$ & time that an artificial blink begins, in units of ms \\
$t_r(j,r)$ & time that a reconstructed saccade begins in an artificial blink,
in units of ms \\
$t_{0r}(j,r)$ & period of a reconstructed saccade in an artificial blink, in
units of ms \\
\hline
\end{tabular}}
\end{center}

\end{widetext}


\begin{thebibliography}{99}

\bibitem{guitton:1991} D. Guitton, R. Simard, and F. Codre,  
Invest. Ophthalmol. Vis. Sci., {\bf 32},
3298 (1991).

\bibitem{engbert:2003} R. Engbert and R. Kliegl,  
 Vision Research, {\bf 43}, 1035 (2003).

\bibitem{kliegl:2006}  R. Kliegl, A. Nuthmann, and R. Engbert,  
Journal of Experimental Psychology: General, {\bf
135}, 12 (2006).

\bibitem{takens} F. Takens, Detecting strange attractors in turbulence in 
{\it Dynamical systems and turbulence} Lecture notes in Math. 898 (Springer,
Berlin 1981).

\bibitem{politi} R. Badii and  A. Politi Complexity, hierarchical structures
and scaling in physics (Cambridge University Press, Cambridge 1997).


\bibitem{paul}  W. Paul and J. Baschnagel, Stochastic Processes, From Physics
to Finance (Springer, Berlin 1999).

\bibitem{ralf} R. Engbert, Progress in brain research, {\bf 154}, 177 (2006).

\bibitem{baptista_PHYSICAA2002} M. S. Baptista and I. L. Caldas, 
Physica A {\bf 312}, 539 (2002).

\bibitem{baptista_plasma} M. S. Baptista, I. L. Caldas, M. V. A. P. Heller,
  and A. A. Ferreira
Phys. of Plasmas {\bf 10}, 1283 (2003).

\bibitem{kitchens} B. P. kitchens, Symbolic Dynamics 
(Springer Verlag, Berlin, 1991).

\bibitem{roland1}  M. S. Baptista, C. Grebogi, and R. K{\"o}berle, 
Phys. Rev. Lett., {\bf 97}, 178102 (2006).

\bibitem{kliegl:2004}  R. Kliegl, E. Grabner, M. Rolfs, and R. Engbert, 
Europ. J. Cog. Psych., {\bf 16}, 262 (2004).

\bibitem{malbouisson:2005} J. M. C. Malbouisson, A. A. V. Cruz, A.
Messias, L. V. O. Leite, and G. D. Rios, 
Invest. Ophthalmol. Vis. Sci., {\bf 46} 857 (2005).

\bibitem{ackman:2005} O. E. Akman, D. S. Broomhead, R. V. Abadi, and
R. A. Clement, 
J. Math. Biol. {\bf 51}, 661 (2005).

\bibitem{comment4} The reason for choosing sequences of length 4, which 
correspond to a time interval of 8ms, is sustained by a reasonable hypothesis
we make. An oscillation period that should be linked to the saccade period has
to happen during a time scale within which relevant dynamical phenomena
happens in the eyes movements. The fastest known microscopic eye
movement happens for time intervals equal or larger than 8ms \cite{ralf}, the
so called microsaccades. This time scale is larger than the time scale within
which possible random fluctuations happen, which we consider to have no direct
link with the period of a saccade that is about to happen or has already
happened.  These fast fluctuations, encoded by symbolic sequences of length
smaller than 4, are relevant to the method being presented. They contribute
largely to the constructions of the digrams. And the digrams are being used
to know if a saccade is about to happen or if it has already happened.

\bibitem{rayner:1998} K. Rayner,  
Psychological Bulletin, {\bf 124}, 372 (1998).

\bibitem{oregan:1987} J. K. O'Regan and A. L\'evy-Schoen, 
Eye movement strategy and tactics in word recognition and reading. 
Attention and Performance, {\bf 2} 363 (1997).

\bibitem{kantz} S. Hallerberg, E. G. Altmann, D. Holstein D, H. Kantz, 
Phys. Rev. E {\bf 75}, 016706 (2007). 

\end{thebibliography}
\end{document}